\documentclass[12pt]{article}
\usepackage{graphicx}
\usepackage{bm}

\usepackage{latexsym}
\usepackage{dcolumn}
\usepackage{epsfig,amssymb,euscript}
\usepackage{amsmath}
\usepackage{amssymb}
\usepackage{array,calc,epsfig}

\usepackage{hyperref} 
\usepackage{chngcntr}
\usepackage[noadjust]{cite}

\usepackage{color}

\usepackage{slashed}

\usepackage{bbold}

\hypersetup{colorlinks=true, linkcolor=blue, urlcolor=blue, citecolor = blue}

\newcommand{\Tr}{{\rm Tr}}

\def\be{\begin{equation}}
\def\ee{\end{equation}}
\def\ba{\begin{eqnarray}}
\def\ea{\end{eqnarray}}

\def\p{\partial}

\def\a{\alpha}
\def\b{\beta}
\def\e{\epsilon}

\def\ff{\phi}

\def\g{\gamma}

\def\d{\delta}

\def\k{\kappa}

\def\m{\mu}

\def\o{\omega}
\def\O{\Omega}

\def\th{\theta}

\def\tt{\tilde}

\def\q{\quad}
\def\we{\wedge}

\def\mf{\mathbf}
\def\mc{\mathcal}

\def\we{\wedge}
\def\dd{\mathrm{d}}

\def\ra{\rightarrow}
\def\lra{\leftrightarrow}

\newcommand{\pr}[1]{\left(#1\right)}

\counterwithin*{equation}{section}

\begin{document}
\baselineskip=15.5pt
\pagestyle{plain}
\setcounter{page}{1}
\newfont{\namefont}{cmr10}
\newfont{\addfont}{cmti7 scaled 1440}
\newfont{\boldmathfont}{cmbx10}
\newfont{\headfontb}{cmbx10 scaled 1728}
\renewcommand{\theequation}{{\rm\thesection.\arabic{equation}}}
\renewcommand{\thefootnote}{\arabic{footnote}}

\vspace{1cm}
\begin{titlepage}
\vskip 2cm
\begin{center}
{\Large{\bf Comments on Holographic Level/Rank Dualities}}
\end{center}

\vskip 10pt
\begin{center}
Riccardo Argurio$^{a}$, Alessio Caddeo$^{b}$
\end{center}
\vskip 10pt
\begin{center}
\vspace{0.2cm}
\textit {$^a$ Physique Th\'eorique et Math\'ematique and International Solvay Institutes, Universit\'e Libre de Bruxelles, C.P. 231, B-1050 Brussels, Belgium}\\
\textit {$^b$ Department of Physics and Instituto de Ciencias y Tecnolog\'ias Espaciales de Asturias (ICTEA)
Universidad de Oviedo, c/ Federico Garc\'ia Lorca 18, E-33007 Oviedo, Spain}\\
\vskip 20pt
{\small{
Riccardo.Argurio@ulb.be, caddeoalessio.uo@uniovi.es}
}

\end{center}

\vspace{25pt}

\begin{center}
 \textbf{Abstract}
\end{center}

\noindent 
We investigate a holographic realization in Type-IIB string theory of pure Chern-Simons theories, and focus on the level/rank dualities that they enjoy.
The level/rank duality is established between the boundary theory, engineered utilizing D3-branes compactified on a $S^1$, and the theory on probe D7-branes located at a specific bulk location. Paying attention to the boundary conditions imposed on the Ramond-Ramond (RR) two- and six-forms, we show that holography precisely reproduces several different level/rank dual pairs. The $SL(2,\mf{Z})$ action relating all these theories is realized through the $SL(2,\mf{Z})$ electric/magnetic duality involving the RR potentials.

\end{titlepage}
\newpage
\tableofcontents
\section{Introduction}
Chern-Simons gauge theories are interesting topological quantum field theories in three-dimensions \cite{Witten:1988hf}. It is of interest to investigate their holographic realization, when the rank of the gauge group is large. This endeavor can shed new light both on the topological theories under scrutiny, and on string theory and holography by analyzing how they reproduce their properties. Chern-Simons theories have been engineered in string theory early on, see for instance \cite{Bergman:1999na,Ohta:1999iv,Maldacena:2001pb} for the non-abelian theories. These set-ups have nevertheless a certain degree of supersymmetry, which makes this description not optimal for pure Chern-Simons theory, even in the vacua where supersymmetry is believed to be broken \cite{Witten:1999ds}.

We will be interested in a particular property of  non-abelian pure Chern-Simons theories, which is level/rank duality \cite{Naculich:1990pa,Mlawer:1990uv,Naculich:2007nc,Hsin:2016blu}. This property was instrumental in understanding some features of the ABJ(M) duality \cite{Aharony:2008ug,Aharony:2008gk} (see also \cite{Aharony:2009fc,Hashimoto:2010bq}), however again in a set-up with supersymmetry and hence extra matter fields coupled to the Chern-Simons theories. A set-up to study pure Chern-Simons theories was devised in \cite{Fujita:2009kw}, using the same trick that allows one to consider holographic duals to pure Yang-Mills theory in three dimensions (as well as in four dimensions \cite{Witten:1998zw}). The starting point is a compactification on a circle $S^1$ in $AdS_5$ of the basic duality between ${\cal N}=4$ Super-Yang Mills and Type-IIB string theory on $AdS_5\times S^5$ with $N$ units of five-form flux \cite{Maldacena:1997re,Aharony:1999ti}. The level of the Chern-Simons theory is provided by $k$ probe D7-branes wrapped on $S^5$ and located at a specific point in the bulk, thus not adding matter to the boundary theory. In the IR, the theory is gapped and we have then a holographic realization of pure Chern-Simons theory.

In \cite{Fujita:2009kw}, it was remarked that holography implements level/rank duality through the duality between the theory that we are supposed to describe at the boundary, with rank related to $N$ and level given by $k$, and the theory that lives on the compactified world-volume of the D7-branes, which has rank related to $k$, and level arising from the flux on $S^5$ and hence given by $N$. The aim of the present note is to flesh out this proposal, in light of the precise rules of level/rank duality that have been spelled out in \cite{Hsin:2016blu}. In particular, since level/rank duality in its simplest incarnation is between $SU$ and $U$ groups, this will involve some non-trivial considerations in holography to distinguish between the two cases, both at the boundary and on the bulk probe branes.

We will analyze the theories by focusing on the couplings to the background fields for their global symmetries, both $U(1)$ zero-form and discrete one-form symmetries. The background field for the continuous symmetry turns out to be (the boundary value of) a component of the RR two-form potential or, crucially, of its dual six-form potential. Going from one formulation to the other involves a bulk electric/magnetic duality, which is holographically interpreted as an $S$ operation on the three-dimensional theory, namely as the gauging of the zero-form global symmetry \cite{Witten:2003ya}. We will also see how to recover the full $SL(2,\mf{Z})$ group action on the theories, and as a consequence the holographic realization of an infinite sequence of level/rank dual theories.

In section \ref{secqft} we review level/rank duality from a purely quantum field theoretic point of view. In section \ref{sechololr} we introduce and then investigate the holographic realization of the duality. We end with a discussion in section \ref{secdisc}.

\section{Quantum field theory perspective}
\label{secqft}
In this section, we provide a review of some properties of the Chern-Simons (CS) theories and the associated level/rank dualities that are worth recalling in view of the discussion of section \ref{sechololr}, where they will be investigated using a holographic approach.

\subsection{$U(1)$ Chern-Simons theories}
\label{subsecabelianqft}
The simplest example of Chern-Simons theory is that with gauge group $U(1)$. Its action reads
\be
S = \frac{k}{4 \pi} \int a \we \dd a \ ,
\ee
where the level $k$ has to be integer so as to make the theory gauge invariant. We use the standard notation $U(1)_k$ to label this theory.
It features a monopole $U(1)$ zero-form global symmetry with current
\be
* J = \frac{\dd a}{2 \pi} \ , \q \q \q \dd * J =0 \ .
\ee
We can couple a background gauge field $A$ to this current,
\be
\label{fqhelagrangian}
S(A) =   \frac{k}{4 \pi} \int a \we \dd a + \frac{1}{2 \pi} \int A \we \dd a  \ .
\ee
This is the action that describes the fractional quantum Hall effect when $A$ is the background electromagnetic field and $a$ the emergent dynamical gauge field. A simple way to find that the level gets fractional is to compute the equation of motion for $a$, 
\be
\label{fqheequationofmotion}
k \dd a = - \dd A   \ ,
\ee
and use it to compute the topological Hall current,
\be
\label{anelianhallcurrent}
* J = \frac{\d S }{\d A} =   \frac{1}{2 \pi } \dd a = - \frac{1}{2 \pi k} \dd A  \ .
\ee
We obtained the topological Hall current by integrating out the dynamical gauge field $a$. In fact, it is not possible to describe the fractional statistics only in terms of the background field $A$. Indeed, in general it is not possible to solve (\ref{fqheequationofmotion}) while respecting the Dirac quantization condition. However, the derivation of the Hall current involves only local properties, thus the equation of motion provides the correct result (see e.g. \cite{Witten:2015aoa}).

Actually, going one step further one can write an effective action for the background field $A$ which reproduces the topological current,
\be
S _{\mathrm{eff}} (A) = - \frac{1}{4 \pi k} \int A \we \dd A \ .
\ee
Notice that to the Lagrangian in (\ref{fqhelagrangian}) we could add a Chern-Simons term of the form $n A \we \dd A/ 4 \pi$ with $n \in \mf{Z}$. Such a \emph{contact term} would shift the current (\ref{anelianhallcurrent}) by a $n \dd A/2 \pi$ term. In fact, the choice of $n$ can be considered as part of the definition of the theory \cite{Witten:2003ya}. However, the observables only depend on the fractional part of the Chern-Simons level \cite{Closset:2012vp}.

Finally, it is worth mentioning that besides the topological $U(1)$ zero-form symmetry, the theory also exhibits a $\mf{Z}_k$ one-form symmetry acting on Wilson lines.

\subsection{$U(N)$ and $SU(N)$ Chern-Simons theories}
\label{subsecnonabelianqft}
Let us consider the case where the gauge group is $U(N)$. The $U(1)$ and the $SU(N)$ factors might admit different levels; we introduce the notation\footnote{The Chern-Simons level depends on the regularization scheme that one adopts. Two commonly used schemes are the Yang-Mills and the dimensional ones. In this paper, we will always adopt the Yang-Mills regularization scheme as in \cite{Hsin:2016blu}. See \cite{Aharony:2015mjs} for a detailed discussion on this point.}
\be
\label{notationgroup}
U(N) _{k,k+Nk'} = \frac{SU(N)_k \times U(1)_{N(k+Nk')}}{\mf{Z}_N} \ , \q \q U(N)_k = U(N)_{k,k} \ .
\ee
The theory exhibits a $U(1)$ monopole zero-form global symmetry. Indeed, calling $b$ the $U(N)$ connection, the current 
\be
* J =  \frac{\dd \Tr b}{2 \pi}
\ee
is identically conserved. We can therefore couple $J$ to a $U(1)$ background connection $C$ and write the action
\be
\label{uennecoupled}
S(C) = \frac{k}{4 \pi} \int \Tr \pr{ b \we \dd b - \frac{2i}{3}b^3} + \frac{k'}{4 \pi} \int \Tr b \we \dd \Tr b + \frac{1}{2 \pi} \int C \we \dd \Tr b \ .
\ee
$SU(N)_k$ Chern-Simons theory can be defined, introducing a $SU(N)$ connection $\hat b$, by
\be
S = \frac{k}{4 \pi} \int \Tr \pr{ \hat b \we \dd \hat b - \frac{2i}{3} \hat b^3}  \ .
\ee
$SU(N)$ gauge theories are expected to admit baryonic configurations with $N$ external quarks \cite{Witten:1998xy}. It is then useful to allow for the possibility of coupling a $U(1)$ background to the baryon current. This can be achieved by working with a $U(N)$ connection $b$ and imposing the constraint 
\be
\label{constrainttraceb}
\Tr b = -  \tilde C \ ,
\ee
where $\tilde C$ is a $U(1)$ background gauge field \cite{Hsin:2016blu}. To make this explicit, let us consider a fermion $\psi$ coupled to the $U(N)$ gauge field $b$,
\be
S = i \int d^3 x  \overline{\psi} \slashed D _b \psi \ .
\ee
Decomposing 
\be
b = \hat b + \frac{1}{N} \Tr b \, \mathbb{1}  
\ee
and using the constraint (\ref{constrainttraceb}) we see that we obtain the coupling 
\be
S_{\tilde C} = \int d^3 x J^\m _B \tilde C_\m \ ,
\ee
where $J^\m _B = \frac{1}{N} \overline{\psi} \g^\m \psi$ is the baryonic current.

The constraint (\ref{constrainttraceb}) can be imposed by introducing an auxiliary $U(1)$ connection $c$ that enters the action as
\be
\label{lagrangemultaction}
S_c = \frac{1}{2 \pi} \int c \we \dd (\Tr b + \tilde C) \ ,
\ee
and over which we integrate in the path integral. Indeed, given a generic three-dimensional manifold $M_3$ and two $U(1)$ connections $\a$ and $\b$, the following result holds \cite{Witten:2003ya}:
\be
\label{fact}
I(\b)=\int D \a \exp \pr{{\frac{i}{2 \pi} \int_{M_3} \a \we \dd \b}} = \d (\b) \ .
\ee
Here, $\d(\b)$ means that $\b$ vanishes up to gauge transformations. Besides enforcing the vanishing of the curvature $\dd \beta$, the sum over all the $U(1)$ bundles annihilates all the holonomies of $\b$. As a result, introducing the action term (\ref{lagrangemultaction}), the integration over $c$ enforces the constraint (\ref{constrainttraceb}).

Hence, we will write the $SU(N)$ Chern-Simons action as\footnote{In fact, if one needs to be careful of the fact that $SU(N)_k$ is a non-spin theory while $U(N)_k$ is such only for even values of $k$, one has to include the following term in the action,
\be
\frac{\e_k}{4 \pi} \int \Tr b \we \dd \Tr b \ , \q \q \q \e_k = 
\begin{cases}
+1 \q \text{if $k$ is odd and positive} \\
-1 \q \text{if $k$ is odd and negative} \\
0 \q \ \ \, \text{if $k$ is even}
\end{cases} \ .
\ee
In the present paper, we are not sensitive to this detail and therefore we will henceforth set effectively $\e_k=0$. In the same vein, we are not going to keep track of terms related to the framing anomaly \cite{Witten:1988hf}.} \cite{Hsin:2016blu}

\be
\label{suennecoupled}
S (\tilde C) = \frac{k}{4 \pi} \int \Tr \pr{ b \we \dd b - \frac{2i}{3}b^3} + \frac{1}{2 \pi} \int c \we \dd ( \Tr b + \tilde C) \ .
\ee
The one-form symmetry of the $SU(N)_k$ theory is $\mf{Z}_N$. On the other hand, recalling (\ref{notationgroup}) and the fact that the one-form symmetry of $U(1)_{N(k+N k ')}$ is $\mf{Z}_{N(k+N k ')}$ we see that the $U(N)_{k,k+Nk'}$ theory displays a $\mf{Z}_{k+Nk'}$ one-form symmetry.

The $SU(N)_k$ Chern-Simons theory is obtained from the $U(N)_k$ theory by gauging the monopole zero-form symmetry. In the language of \cite{Witten:2003ya}, this is the $S$ operation. We can represent it by substituting the background gauge field $C$ with a dynamical one $c$ and coupling the topological current of the latter to a new background field $\tilde C$, namely 
\be
S (C) \ra S(c) + \frac{1}{2 \pi} \int c \we \dd \tilde C \ .
\ee
In three dimensions,  after gauging the $U(1)$ zero-form global symmetry of a theory we find another theory with a $U(1)$ zero-form global symmetry. In our example, we started with the $U(N)_k$ theory and we gauged its monopole symmetry in order to obtain the $SU(N)_k$. This latter theory exhibits a baryon zero-form global symmetry, with current 
\be
* J = \frac{\dd \Tr c}{2 \pi} \ .
\ee
Similarly, applying $S$ to the $SU(N)_k$ theory one finds the $U(N)_k$. 
Besides the $S$ operation, another operation, called $T$, is defined by adding a contact term for the background gauge field coupled to the $U(1)$ zero-form global symmetry,
\be
S(C) \ra S(C) + \frac{1}{4 \pi} \int C \we \dd C \ .
\ee
As we mentioned in section \ref{subsecabelianqft}, the addition of such a contact term for the background gauge field is in itself a trivial operation since it does not affect the theory's observables. However, it is not entirely trivial because it does not commute with the $S$ operation. Indeed, one can easily show that $S$ and $T$ generate the $SL(2,\mf{Z})$ group \cite{Witten:2003ya}.

\subsection{Level/rank dualities}
\label{commentonlevelranksimilar}
The theories that we have introduced so far are not inequivalent. Indeed, they are related by known (exact) \emph{level/rank dualities} \cite{Naculich:1990pa,Mlawer:1990uv,Naculich:2007nc,Hsin:2016blu}. Let us write the simplest ones,
\begin{subequations}\label{lrdualities}
\ba
\label{sunuklevelrankduality}
SU(N)_k &\lra&  U(k)_{-N} \ , \\
\label{uulevelrankduality}
U(N)_{k,k\pm N} &\lra& U(k)_{-N,-N\mp k}  \ .
\ea
\end{subequations}
We will not review the proof of these dualities. Rather, we show that the topological (Hall) currents match between the two sides of these dualities. This is the observable that we will be able to compute within the holographic context and will allow us to identify the involved Chern-Simons theories.

Let us start with the $SU(N)_k$ theory (\ref{suennecoupled}). 
Let us separate the terms involving $\Tr b$ from the remaining ones,
\be\label{actionsun}
S(\tilde C) = S_{\text{traceless}} +  \frac{k}{4 \pi N} \int \Tr  b \we \dd \Tr b + \frac{1}{2 \pi} \int c \we \dd  \Tr b + \frac{1}{2 \pi} \int \tilde C \we \dd c \ .
\ee
Taking the equations of motion for $\Tr b$ and $c$, we find
\begin{subequations}
\label{sunoddeqmotion}
\ba
 \frac{k }{N} \dd \Tr b &=&-  \dd c  \ , \\
\dd \Tr b &=& -\dd \tilde C  \ .
\ea
\end{subequations}
The topological current is then
\be
\label{currentsuenne}
* J = \frac{\d S}{\d \tilde C} = \frac{1}{2 \pi } \dd c =  \frac{k}{2 \pi N} \dd \tilde C \ .
\ee
The effective action reproducing the topological current \eqref{currentsuenne} is
\begin{align}\label{Leffsun}
S_{\mathrm{eff}}(\tilde C)=\frac{k}{4\pi N} \int  \tilde C \we \dd \tilde C\ .
\end{align}

Let us perform the same computation starting from the $U(N)_{k,k+Nk'}$ theory. After separating the trace part from the other components, the action (\ref{uennecoupled}) reads
\be
S(C) = S_{\text{traceless}}  + \frac{k + Nk'}{4 \pi N} \int  \Tr b \we \dd \Tr b + \frac{1}{2 \pi} \int  C \we \dd \Tr b \ .
\ee
The equation of motion for $\Tr b$ is
\be
\frac{k + Nk'}{N}  \dd \Tr b =- \dd C  \ ,
\ee
so that the topological current reads
\be
\label{topologicalcurrentunkk'}
* J = \frac{\d S}{\d C} = \frac{1}{2 \pi } \dd \Tr b =  - \frac{N}{2 \pi (k+Nk')} \dd C \ .
\ee
The case $k'=0$ gives the $U(N)_k$ theory.  
Again, we can write an effective action for $C$ also in this case. It reads
\begin{align}
\label{Leffun}
S_{\mathrm{eff}}(C)=-\frac{N}{4\pi k} \int C \we \dd C \ .
\end{align}
Confronting the above expression with \eqref{Leffsun}, we see that the topological current of the $U(k)_{-N}$ theory coincides with that of the $SU(N)_k$ theory, as expected from the level/rank duality (\ref{sunuklevelrankduality}). On the other hand, it is important to notice that the effective action for $C$ for generic $N$ and $k$ is different depending on whether we are considering an $SU(N)_k$ or a $U(N)_k$ CS theory.

For $k' =\pm 1$ we have the $U(N)_{k,k\pm N}$ theory, with current 
\be
\label{topologicalcurrentunkk'one}
* J = \frac{\d S}{\d C} = \frac{1}{2 \pi } \dd \Tr b =  - \frac{N}{2 \pi (k\pm N)} \dd C \ .
\ee
As we wrote in (\ref{uulevelrankduality}), this theory is known to be level/rank dual to the $U(k)_{-N,-N\mp k}$ one \cite{Hsin:2016blu}. Let us show that the currents indeed match. Following the same steps as before, we write its action as
\be
S(C) = S_{\text{traceless}}  - \frac{N\pm k}{4 \pi k} \int \Tr b \we \dd \Tr b + \frac{1}{2 \pi} \int C \we \dd \Tr b \ .
\ee
The equations of motion read
\be
\frac{ N\pm k}{k}  \dd \Tr b = \dd C  \ ,
\ee
and therefore the current is 
\be
* J = \frac{\d S}{\d C} = \frac{1}{2 \pi } \dd \Tr b =   \pm \frac{k}{2 \pi (k\pm N)} \dd C \ .
\ee
We see that the two currents match up to $\pm \dd C/2 \pi$. As we mentioned at the end of section \ref{subsecabelianqft}, such a contribution comes from a contact term in the Lagrangian, which does not affect the observables of the theory.  Thus, the level/rank duality (\ref{uulevelrankduality}) is an example in which there is a non-trivial map between contact terms.

A last observation is the following. If in three dimensions a duality between two theories with a $U(1)$ zero-form symmetry is established, it is clear that acting on both sides with the same $SL(2, \mf{Z})$ operation, one finds another duality. Indeed, it is possible to show \cite{Hsin:2016blu} that the level/rank dualities \eqref{uulevelrankduality} are obtained by acting on the pair \eqref{sunuklevelrankduality} with $T^{\pm1}$ and then with $S$, on both sides. Taking different combinations of $T$ and $S$ one then finds an infinite set of level/rank dualities relating Chern-Simons theories, although in some cases the latter can be defined only formally \cite{Choi:2019eyl}.

\section{Holographic level/rank dualities}
\label{sechololr}
In this section, we describe how holography reproduces the level/rank dualities for the cases discussed in section \ref{secqft}. We start by introducing the holographic model and then we investigate the different cases.

\subsection{Holographic set-up: the cast of characters}
\label{holosetup}
We start by recalling the holographic set-up of \cite{Fujita:2009kw}. It aims at finding the (top-down) gravity dual of a three-dimensional pure $SU(N)$ or $U(N)$ gauge theory with a level $k$ Chern-Simons term.
The construction is the following.

Let us consider a stack of $N$ coincident D3-branes which wrap an $S^1$ of length $L$ with antiperiodic (periodic) boundary conditions for fermions (bosons), so as to explicitly break supersymmetry. In this way, adjoint fermions and scalars take masses and at distances greater than $L$ we are left with $2+1d$ gauge fields. The theory on the D3-branes admits a coupling with the axion field $C_0$,
\be
\label{d3actionc0}
S_{C_0} = \frac{\m_3}{2} \int_{\mf{R}^{1,2} \times S^1} C_0 \Tr \pr{ 2 \pi f  \we 2 \pi f } = \frac{1}{4 \pi } \int_{\mf{R}^{1,2} \times S^1} C_0 \dd \Tr \pr{a \we \dd a - \frac{2i}{3} a^3} \ .
\ee
Here, $f$ is the field strength of the $U(N)$ connection $a$ and
\be
\m_p = \frac{1}{(2 \pi)^p} \ ,
\ee
where we take units where $l_s=1$.

Let us take an axion field of the form
\be
\label{ansatzC0}
C_0 (x_3) =  \frac{ k x_3}{L} \ ,
\ee
where $x_3$ is the coordinate along $S^1$. In this way\footnote{ We take the volume form to be ordered according to $\dd x^0 \we \dd x^1 \we \dd x^2 \we \dd x^3 \we \dd r \we \o_5$. Equivalently, we take $\e_{0123 r\th_1 \dots \th_5} = + 1$. In particular, when performing the integral over $dx^3$, we bring the differential to the right of $\dd x^0 \we \dd x^1 \we \dd x^2$. This fixes the sign of $S_{C_0}$.}
\be
S_{C_0} = \frac{k}{4 \pi} \int_{\mf{R}^{1,2}} \Tr \pr{a \we \dd a - \frac{2i}{3} a^3} \ .
\ee
As a result, we find a $2+1d$ Yang-Mills theory with gauge group $U(N)$ and a level $k$ Chern-Simons term, that is the theory $U(N)_k$.
At low energies the Yang-Mills term (coming from the Dirac-Born-Infeld piece of the D-brane action, that we do not write) is irrelevant so that the theory flows into pure $U(N)_k$ Chern-Simons theory.\footnote{In order to obtain the low-energy theory, we have integrated out the adjoint fermions. In principle, one could expect a shift of the Chern-Simons level due to this integration. This is not the case. Indeed, let us take one of the $4d$ Weyl fermions and let us compactify it on the circle $S^1$, 
\be
\psi (x^\m, x_3) = \sum _n \psi _n (x^\m) e^{2 \pi i n x_3/L} \ , \q \q  n = \pm 1/2, \pm 3/2 \dots \ ,
\ee
where the condition on $n$ comes from the antiperiodicity of the fermion. A $4d$ Weyl fermion $\psi$ gives a tower of $3d$ Dirac fermions $\psi_n$ with mass $m \propto n/L$. We thus see that there are no zero modes; moreover, massive Dirac fermions come in pairs with the same mass but opposite sign. As a result, there is no shift in the Chern-Simons level.} Let us mention that since we will take $k$ to be finite and $N$ to be large, we will be in the regime where $N/k\gg1$ and so the Yang-Mills-Chern-Simons theory is strongly coupled. The theory is gapped, with gap proportional to $N$, but does not confine. Hence the (topologically non-trivial) theory below the gap can be safely described by the pure Chern-Simons theory.

Note that, until now, we are considering the theory on the D3-branes as $U(N)$ and not $SU(N)$. It is customary in holography to discard the overall $U(1)$ since at the Yang-Mills Lagrangian level, it decouples completely from the rest of the dynamics. It is associated to the center-of-mass of the stack of D3-branes, whose location is arbitrary in the configuration one starts with. However, in the present set-up where we crucially have a Chern-Simons level, such decoupling is no longer granted, in the sense that the presence or not of the overall $U(1)$ gauge factor affects the topological theory at low energies. We will thus have to be careful in building the correct holographic theory for $SU(N)_k$. As we will see, we can use $U(N)_k$ as a starting point.

The $U(N)$ theory features an identically conserved $U(1)$ current,
\be
\label{flatholohallcurrent}
* J =  \frac{\dd \Tr  a}{2 \pi} \ .
\ee
This current is coupled to the RR two-form $C_2$ through the Wess-Zumino coupling. Fixing the Kalb-Ramond field $B_2 =0$, we have
\be
\label{couplingC2D3branes}
S_{C_2} = \m_3 \int_{\mf{R}^{1,2} \times S^1} C_2 \we 2 \pi \Tr  f \ .
\ee
Taking the ansatz 
\be
\label{ansatzC2}
C_2 = 2 \pi C \we \dd x^3 / L 
\ee
and integrating over $S^1$, we find 
\be
\label{d3couplingtoC}
S_{C_2} = \frac{1}{2 \pi} \int_{\mf{R}^{1,2}} C \we \dd  \Tr \, a \ ,
\ee
which is the coupling of the $U(1)$ current $J$ with a classical background field $C$ in $2+1d$. 

We have thus learned that a signature of the fact that we are describing a $U(N)$ gauge theory is that we can see the coupling of its topological $U(1)$ current to a background field, which is a particular component of the RR field $C_2$. This observation is the key to being able to describe also an $SU(N)$ gauge theory. Indeed, as reviewed in section \ref{secqft}, in $3d$ one obtains the $SU(N)$ gauge theory from the $U(N)$ by gauging its topological symmetry. In holography, gauging a global symmetry amounts to going to the alternative quantization for the bulk gauge field dual to the global current. This in turn is achieved in the simplest way by dualizing the bulk gauge field. For us, it will amount to formulate the bulk dynamics in terms of the $C_6$ RR potential instead of $C_2$.  

However first let us consider the backreaction of the present configuration, in other words the holographic dual theory. In the absence of a CS level, the near-horizon limit of the D3-branes geometry reads
\begin{subequations}
\label{gravitybackground}
\be
ds^2 = \frac{r^2}{R^2} \pr{ dx_\m dx^\m + f(r) dx_3 ^2} + \frac{R^2}{r^2} \frac{dr^2 }{f(r)} + R^2 d \O_5 \ , 
\ee
with
\begin{align}
\label{dilatonfi}
e^\ff&=g_s  \ ,           &  f(r) &= 1- \frac{r_0 ^4}{r^4} \ ,           \\
\label{rparameter}
F_5 &=\dd C_4 = - {(2 \pi)^4 N} ( \o _5 +* \o_5) \ ,         &  R^4&=4 \pi g_s N l_s ^4 \ ,
\end{align}
\end{subequations}
where $\omega_5$ is the $S^5$ (unit) volume form.\footnote{The sign of the expression for $F_5$ is consistent with the sign of the D3-brane action \eqref{d3actionc0}, in the conventions that we will detail below.}
The topology of the cigar-shaped $(x_3,r)$ subspace is that of the disk $D_2$, whose boundary at $r=r_\infty$ is $S^1$ with coordinate $x_3$. 

Contrarily to the initial situation, where the $S^1$ described a cylinder and thus the axion winding could be purely topological, after backreaction the $(x_3,r)$ subspace is simply-connected and therefore in order to have a $C_0$ which is multi-valued along $S^1$ there must be a defect. Indeed, according to the holographic prescription where the D3-branes can be thought to live at the boundary of the asymptotically $AdS$ spacetime, we need to recover \eqref{ansatzC0} at $r=r_\infty$. Calling $F_1=\dd C_0$, this entails\footnote{The choice of order in the volume form implies sometimes a minus signs in the Stokes theorem. Indeed, taking for simplicity a form $A$ that only depends on $r$, we have
	\be
	\int_{D_2} \dd A = \int_{D_2} \p_r A_{x^3} \dd r \we \dd x^3 \propto - \int_{S^1} A|_{S^1} \ .
	\ee
The minus sign comes from the fact that according to $\e_{0123 r\th_1 \dots \th_5} = + 1$, we have to order the differentials as $\dd x^3\we \dd r$ before performing the integral.}
\be
\label{bianchiidentityF1}
\int _{S^1} F_1 = - \int _{D_2} \dd F_1 =   k\ .
\ee
A defect is therefore needed in order to violate the Bianchi identity for $F_1$. Such a defect is provided by $k$ D7-branes wrapped on $S^5$ and pointlike in the $(x_3,r)$ subspace, i.e.~they can be thought as being located at the origin of the disk $D_2$. Being localized at the tip of the cigar, i.e.~in the deep bulk, they do not alter the matter content of the UV theory. 
Note that \eqref{ansatzC0}, thought as valid for any value of $r$ (except $r_0$ where $x_3$ is not well-defined), solves the equation of motion for $C_0$. As long as we keep $k$ fixed in the large-$N$ limit, $C_0$ does not backreact on the geometry, and consistently also the D7-branes will be considered as non-backreacting probes. 

Since we have now established that the gravity dual of the gauge theory with non-trivial CS level has to incorporate probe D7-branes, we have to take into account their world-volume action when considering the bulk equations of motion. 

Before writing the latter, recall that as much as the closed string modes on the cigar are known to be gapped, the fluctuations of the D7-branes are also gapped.  
Indeed, calling $b$ the $U(k)$ connection of the D7-brane theory and $\mc{F}$ its field strength, we have from a piece of the Wess-Zumino action\footnote{Again, the sign of the following action is fixed by the sign of the expressions \eqref{bianchiidentityF1}.}
\be
\frac{\m_7}{2} \int_{\mf{R}^{1,2} \times S^5} C_4 \we  \Tr \pr{ 2 \pi \mc{F}  \we 2 \pi  \mc{F} } = - \frac{N}{4 \pi} \int_{\mf{R}^{1,2}} \Tr \pr{ b \we \dd b - \frac{2i}{3} b^3} \ ,
\ee
where we used that $F_5 = \dd C_4$ satisfies 
\be
\int_{S^5} F_5  = - (2 \pi)^4 N \ .  
\ee
This Chern-Simons term controls the physics at low energies (in particular, the Yang-Mills term from the DBI piece of the action becomes irrelevant) and as a result, in the world-volume of the D7-brane lives a three-dimensional topological quantum field theory.

The bulk holographic solution thus provides us naturally with a $U(k)_{-N}$ Chern-Simons theory. Note that here the fact that we are in the regime $N/k\gg1$ means that this Yang-Mills-Chern-Simons theory is weakly coupled, it becomes topologically gapped while still in the perturbative regime. Again, the $U(1)$ gauge factor related to the center-of-mass of the D7-branes does not decouple straightforwardly.

In this respect, another term in the D7-brane Wess-Zumino action is important.
The $U(k)$ gauge theory on the D7-branes features a topological symmetry with current
\be
* \tilde J =  \frac{\Tr \mc{F}}{2 \pi} \ .
\ee
Such a current couples to the $C_6$ Ramond-Ramond field,
\be
\label{d7couplingc6}
S_{C_6} = 2 \pi  \m_7 \int_{\mf{R}^{1,2} \times S^5} C_6 \we  \Tr \mc{F} \ .
\ee
If we use the ansatz
\be
\label{ansatzC6}
C_6  = (2 \pi)^5 \tilde C \we \omega_5 \ ,
\ee
we find the three-dimensional coupling to the external gauge field $\tilde C$,
\be
\label{d7couplingCtilde}
S_{C_6} = \frac{1}{2 \pi} \int_{\mf{R}^{1,2}} \tilde C \we \Tr \mc{F} \ .
\ee
We thus learn that the coupling of the topological symmetry current of the $U(k)$ theory can be naturally written when the $C_6$ RR potential is used to describe the bulk dynamics, i.e.~in the dual formulation with respect to \eqref{d3couplingtoC}.

Note that \eqref{d7couplingCtilde} is best understood as being evaluated at $r=r_0$, where the D7-branes lie. The complete action of the D7-branes' theory is therefore
\be
S_{D7} = - \frac{N}{4 \pi} \int_{\mf{R}^{1,2}} \Tr \pr{ b \we \dd b - \frac{2i}{3} b^3} + \frac{1}{2 \pi} \int_{\mf{R}^{1,2}} \tilde C \we \Tr \mc{F} \ .
\ee
Integrating over $b$ (with the caution advocated in section \ref{subsecabelianqft}) we can write the corresponding effective action
\be
\label{d7effectiveaction}
S_{D7} =   \frac{k}{4\pi N} \left. \int_{\mf{R}^{1,2}} \tilde C\wedge \dd \tilde C\right|_{r=r_0} \ .
\ee
For later convenience, we explicitly wrote that here $\tilde C$ is evaluated at $r=r_0$.

It is important to notice that the background fields $C$ and $\tilde C$ are not identified. They are however related in that they are particular components of $C_2$ and $C_6$, respectively, which in turn are dual RR potentials in Type-IIB supergravity. This relation is crucial for our arguments in the following.

To summarize, we started with a flat spacetime in which $N$ D3-branes are wrapped on a cylinder with $k$ units of non-trivial $C_0$ monodromy, i.e.~$F_1$ flux on the $S^1$. After backreaction, we ended up with a curved spacetime, including an $S^5$ with $N$ units of $F_5$ flux and a cigar-shaped factor with $k$ D7-branes at the tip. In both situations at low energies we have a three-dimensional topological Chern-Simons theory, and we are going to argue that it is a level/rank dual pair.

\subsection{Effective action for the background field}
\label{sechallonlysugra}
We are now going to use the equations of motion of the bulk supergravity fields, coupled to the probe D7-brane action, in order to recover an effective action for the background fields $C$ or $\tilde C$. This will be a way to reproduce holographically the results of section \ref{commentonlevelranksimilar}, and to see more precisely how the level/rank dual pairs \eqref{lrdualities} are achieved.

The action of Type-IIB supergravity reads
\be
\label{typeIIbaction}
S_{\text{IIB}} = S_{\text{NS}} + S_{\text{RR}} + S_{\text{CS}} \ ,
\ee
where 
\begin{subequations}
\label{azionesugradueb}
\ba
S_{\text{NS}} &=& \frac{1}{2 \k_{10}^2} \int d^{10 } x \sqrt{-G} \, e^{-2 \ff} \pr{\mc{R} + 4 \p_M \ff \p^M \ff - \frac{1}{2} |H_3|^2 }  \ , \\
S_{\text{RR}} &=& - \frac{1}{4 \k_{10}^2} \int d^{10 } x \sqrt{-G} \pr{|F_1|^2 + |\tt{F}_3|^2 +  \frac{1}{2} |\tt{F}_5|^2 }  \ , \\
S_{\text{CS}} &=& - \frac{1}{4 \k_{10} ^2} \int B_2 \we F_3 \we F_5 \ .
\ea
\end{subequations}
Here, $2 \k_{10} ^2 = (2 \pi)^7$, we defined
\begin{subequations}
\ba
\tt{F}_3 &=& F_3 - C_0 H_3 \ ,  \\
 \tt{F}_5 &=& F_5 + \frac{1}{2} B_2 \we F_3 - \frac{1}{2} C_2 \we H_3 \ ,
\ea
\end{subequations}
and it is understood that $\tilde F_5=*\tilde F_5$ is imposed after the equations of motion have been derived from the above action.\footnote{According to our conventions,
\be
 \int d^{10 } x \sqrt{-G}  |F_p|^2  =\frac{1}{p!} \int d^{10} x \sqrt{|G|} F_{\m_1 \dots \m_p} F^{\m_1 \dots \m_p} = \int F_p \we * F_p\ .
\ee}

A central role is played by the topological term of the action, which dominates for large values of the holographic coordinate.
It comes from $S_{\text{CS}}$ and from the $|\tilde F_3|^2$ and $|\tilde F_5|^2$ contributions to $S_{\text{RR}}$. Using the ans\"atze (\ref{ansatzC2}) and (\ref{ansatzC6}), it reads
\be
\label{topologicalsugraterm}
S_{\text{top}} =  - \frac{1}{(2 \pi)^2} \int_{\mf{R}^{1,2} \times r} \pr{ N B_2 \we \dd C - k B_2 \we \dd \tilde C} \ .
\ee
Note that we are using above a mixed notation where both $C$ and $\tilde C$ are present. Of course we have to keep in mind that they are not independent but one the dual of the other.
As we will discuss, this term plays an important role when imposing the boundary conditions on the RR fields.

To the supergravity action for the bulk fields, we must add the following action for the probe D7-branes:
\begin{align}
S_{D7}  =   \m_7 \sum_q \int_{\mf{R}^{1,2} \times S^5} C_{q} \we \Tr \exp \pr{2 \pi \mc{F} - B_2} \ ,
\end{align}
where we have neglected the DBI part of the action. We remind that $B_2$, $C_q$ are actually the pullback on the D7-brane world-volume of these quantities. For our purposes, the relevant terms of the action will be the following:
\begin{align}\label{D7wzaction}
S_{D7}  =   \mu_7 \int \left[k\ C_8 + C_6\wedge \Tr ({2\pi{\cal F}} -   B_2)+\frac12C_4 \wedge \Tr ({2\pi{\cal F}} - B_2)^2 \right] \wedge \delta_2 \ ,
\end{align}
where we have also taken the opportunity to write the action as an integral over all spacetime, localized at the tip of the cigar/center of the disk $D_2$ by the (closed) two-form $\delta_2$, which can thus be seen as the delta-function source for the D7-branes.\footnote{Consistenly with our conventions, the volume form of the disk $D_2$ is defined with the sign such that $\d_2 \propto  \dd x^3 \we \dd r$.}

We will consider a fixed background where the metric, dilaton and $F_5$ are given by \eqref{gravitybackground}. On top of this, we will consider a non-backreacting $F_1$ flux, and fluctuations of $C_2/C_6$ and ${\cal F}$. Crucially, we also consider the equations of motion for $B_2$, though we consistently set the profile of $B_2$ eventually to zero in all the equations. 

The equations of motion (or Bianchi identities) that we get are the following. 
For $C_0/C_8$, we have 
\begin{align}
\dd *F_1=0 \ , \qquad \qquad 
\dd F_1 = - k \delta_2\ .
\end{align}
The Bianchi identity translates the fact that there are $k$ D7-branes in the bulk, more specifically the source is given by the first term in \eqref{D7wzaction}. For $C_2/C_6$, we have
\begin{align}
\label{eqsmotionf3}
\dd *F_3=0\ , \qquad \qquad \dd F_3= - 2 \pi \Tr {\cal F}\wedge \delta_2\ .
\end{align}
Again, the Bianchi identity (or equation of motion for $F_7=*F_3$) has a non-trivial source due to the second term in \eqref{D7wzaction}.

The equation of motion for $B_2$ imposes the following constraint:
\begin{align}
\dd C_0 \we * F_3 + C_0 \we \dd * F_3  + F_3 \we F_5 = - (k\ C_6 + C_4 \wedge 2 \pi\Tr {\cal F})\wedge \delta_2\ .
\end{align}
Using the equations of motion for $F_3$, the above expression can be simplified to 
\begin{align}
\label{eomB2}
F_1\wedge F_7 + F_3\wedge F_5 = - (k\ C_6 + C_4 \wedge 2 \pi \Tr {\cal F})\wedge \delta_2\ .
\end{align}
Finally, the equations of motion for the gauge field $b$ on the D7-branes are
\be
\label{eombd7}
\pr{ F_7 + \frac{2 \pi}{k} F_5\wedge \Tr {\cal F}} \wedge \delta_2 = 0 \ .
\ee
In particular, from this equation we derive a simple relation between the pullback of the curvature of $\tilde C$ and the gauge field on the $D7$-branes,
\be
\label{relationfortildeCatthetip}
\left. \dd \tilde C \right|_{r=r_0}= \frac{N}{k} \Tr {\cal F}    \ .  
\ee

We solve the Bianchi identity for $C_0$ by taking
\begin{align}
F_1=  k \frac{\dd x^3}{L}\ ,
\end{align} 
and recall that (omitting its self-dual part along $\mf{R}^{1,2}\times D_2$)
\begin{align}
F_5 = - (2\pi)^4 N \omega_5\ .
\end{align}
For $r>r_0$, \eqref{eomB2} reads then
\begin{align}
\label{onshellconstraint}
k \frac{\dd x^3}{L}\wedge F_7 =  (2 \pi)^4 N F_3\wedge \omega_5\ ,
\end{align}
which in turn implies
\begin{align}
\dd \tilde C|_{r>r_0}= \frac{N}{k}  \dd C|_{r>r_0}  \ .
\end{align}
Moreover, integrating the equations (\ref{eomB2}) over $D_2 \times S^5$ we find
\ba
\label{relationforCatthetip}
 C|_{r=r_0} =  \Tr b   \ , 
\ea
so that using \eqref{relationfortildeCatthetip} we also have
\ba
\label{CCTildetip}
   \dd \tilde C |_{r=r_0}= \frac{N}{k}  \dd C|_{r=r_0}  \ , 
\ea
i.e.~the on-shell relation between $C$ and $\tilde C$ is valid everywhere.

We now want to evaluate the bulk action on-shell. Actually, our aim is to find an effective action for either $C$ or $\tilde C$. However, part of the bulk solution consists of the D7-branes, with the world-volume action for the dynamical field $b$. We have already seen that such action yields a $U(k)_{-N}$ Chern-Simons theory with a coupling to the $\tilde C$ component of $C_6$. As we write in (\ref{d7effectiveaction}), this results in an effective action for $\tilde C$, though it would be localized at the tip of the cigar $r=r_0$, which is inconvenient for holography, where source fields are supposed to be `measured' at the boundary $r=r_\infty$. Nevertheless, let us see if the bulk supergravity action allows us to extract an effective action for $C$ or $\tilde C$. We will do this by imposing different boundary conditions on the RR fields.

\subsubsection{$SU(N)_k$ theory}
\label{secholosu(n)}
As we have just mentioned, the on-shell D7-brane action is formulated naturally in terms of $\tilde C$.  We thus start by imposing Dirichlet boundary conditions on $C_6$, that is on $\tilde C$, so that the whole on-shell action should depend on it. 

As we anticipated, the topological term (\ref{topologicalsugraterm}) in the supergravity action plays a crucial role. Indeed, since we fix the boundary value of $\tilde C$, we cannot at the same time fix the boundary value of $C$. As a result, from the first term in (\ref{topologicalsugraterm}) we read that $B_2$ has to be a holonomy taking values in $\mf{Z}_N$, so the dual quantum field theory displays a $\mf{Z}_N$ one-form symmetry. (For these arguments, see \cite{Aharony:1998qu,Witten:1998wy,Hofman:2017vwr} and in particular \cite{Bergman:2020ifi} for a very similar situation.) As we mentioned in section \ref{subsecnonabelianqft}, this is the one-form symmetry of the $SU(N)_k$ theory. We now show that holography consistently yields the effective action of the $SU(N)_k$ theory.

Since we want to impose Dirichlet boundary conditions on $C_6$, it is convenient to dualize the Type-IIB supergravity action to write it in terms of $C_6$ rather than $C_2$. The only relevant part of the supergravity action is the one with $F_3$, that we rewrite
\begin{align}
S_{\text{IIB}}= - \frac{1}{4 \k_{10}^2} \int_{\mf{R}^{1,2} \times D_2\times S^5}  F_7\wedge *F_7 \ .
\end{align}
Since $\dd*F_3=\dd F_7=0$ everywhere, we can write $F_7=\dd C_6$ in the above action and integrate by parts. We obtain
\be
S_{\text{IIB}}= - \frac{1}{4 \k_{10}^2} \int_{\mf{R}^{1,2}\times S^1 \times S^5} C_6\wedge F_3 + \frac{1}{4 \k_{10}^2} \int_{\mf{R}^{1,2} \times D_2 \times S^5} C_6 \wedge \dd F_3\ .
\ee
We now consider the ansatz (\ref{ansatzC6}) and the equation (\ref{eqsmotionf3}), to get
\be
S_{\text{IIB}}= - \frac{(2 \pi)^5}{4 \k_{10}^2} \int_{\mf{R}^{1,2}\times S^1 \times S^5} \tilde C\wedge \omega_5\wedge F_3 -  \frac{(2 \pi)^6}{4 \k_{10}^2} \int_{\mf{R}^{1,2} } \tilde C \wedge \Tr {\cal F}\ ,
\ee
where in the second term we have performed the integral over $D_2 \times S^5$.
Note that the first term is at the boundary, namely $r=r_\infty$, while the second term is at $r=r_0$.
For the first term we use \eqref{onshellconstraint} and the ansatz (\ref{ansatzC6}) for $C_6$ to write
\begin{align}
S_{\text{IIB}}=   \frac{k}{4\pi N}\left. \int_{\mf{R}^{1,2}} \tilde C\wedge \dd \tilde C\right|_{r=r_\infty}  -  \frac{1}{4 \pi } \left. \int_{\mf{R}^{1,2} } \tilde C \wedge \Tr {\cal F}\right|_{r=r_0} \ .
\end{align} 
The last step is then obtained by using the equations of motion for $b$, namely \eqref{relationfortildeCatthetip}, so that we finally get
\begin{align}
S_{\text{IIB}}=  \frac{k}{4\pi N}\left. \int_{\mf{R}^{1,2}} \tilde C\wedge \dd \tilde C\right|_{r=r_\infty}  -  \frac{k}{4\pi N} \left. \int_{\mf{R}^{1,2}} \tilde C\wedge \dd \tilde C\right|_{r=r_0}\ .
\end{align}

The on-shell supergravity action exhibits two terms, one evaluated at the spacetime boundary as customary in holography, and one evaluated at the center of the disk $D_2$ where the D7-branes lie. The holographic prescriptions demand the latter to be canceled in the final result. Indeed, taking into account also the on-shell D7-brane action (\ref{d7effectiveaction}), we find such cancellation:
\begin{align}
\label{holosuneffectiveaction}
S_{\text{IIB}} + S_{D7} =  \frac{k}{4\pi N}\left. \int_{\mf{R}^{1,2}} \tilde C\wedge \dd \tilde C\right|_{r=r_\infty}\ .
\end{align}
We have thus found an effective action for $\tilde C$, with the latter acting as background gauge field at the boundary of the holographic spacetime. Recalling (\ref{Leffsun}), we recognize (\ref{holosuneffectiveaction}) to be the effective action of the $SU(N)_k$ Chern-Simons theory. The computation confirms that the bulk theory supplemented with Dirichlet boundary conditions for $\tilde C$ is dual to the $SU(N)_k$ theory, as anticipated at the beginning of this subsection by the argument based on the one-form symmetry. 

Furthermore, in section \ref{holosetup}, the theory on the D7-branes was precisely identified with the $U(k)_{-N}$ Chern-Simons one. We therefore find that imposing Dirichlet boundary conditions on $\tilde C$, the holographic duality precisely reproduces the $SU(N)_k \lra U(k)_{-N}$ level/rank duality. In the following, we will impose different boundary conditions in order to find other level/rank dualities.

\subsubsection{$U(N)_{k}$ theory}
\label{subsecunk}
Another simple case is that in which we impose Dirichlet boundary conditions on $C_2$, i.e.~on $C$. This is equivalent to imposing Neumann boundary conditions on $C_6$ (i.e.~on $\tilde C$). It means that the boundary value of $C_6$ is now free to fluctuate, so it acts as a dynamical field rather than a background one. In other words, we are gauging the symmetry to which $C_6$ was coupled. 

We are thus gauging the baryonic zero-form symmetry of the $SU(N)_k$ theory. As we discussed in section \ref{subsecnonabelianqft}, this produces a $U(1)$ zero-form symmetry, now of a monopole kind. As a result, in the present subsection we expect to find the holographic dual of the $U(N)_k$ theory.

The appearance of the $U(N)_k$ theory is corroborated also by the symmetry argument made at the beginning of subsection \ref{secholosu(n)}. Let us look again at the topological action (\ref{topologicalsugraterm}). Since we fix the boundary value of $ C$, from the second term in (\ref{topologicalsugraterm}) we read that $B_2$ has to be a holonomy taking values in $\mf{Z}_k$, so the dual quantum field theory displays a $\mf{Z}_k$ one-form symmetry. As we mentioned in section \ref{subsecnonabelianqft}, this is the one-form symmetry of the $U(N)_k$ theory.

Since we are imposing Dirichlet boundary conditions on the RR field $C_2$, in this subsection it is convenient to work with the latter. The relevant action term is
\be\label{iibactionf3}
S_{\text{IIB}} = - \frac{1}{4 \kappa_{10} ^2} \int_{\mf{R}^{1,2} \times D_2\times S^5} F_3 \we * F_3 \ .
\ee
Since $F_3$ is exact everywhere but at $r=r_0$, let us write
\be
\label{formofF3}
F_3 = \dd C_2 - 2 \pi \Tr b \we \d_2 \ ,
\ee
which solves (\ref{eqsmotionf3}). We then have
\ba
S_{\text{IIB}} = - \frac{1}{4 \kappa_{10} ^2} \int_{\mf{R}^{1,2} \times S^1 \times S^5} C_2 \we * F_3  + \frac{2 \pi}{4 \kappa_{10} ^2} \int_{\mf{R}^{1,2} \times D_2\times S^5} \Tr b \we \d_2 \we *F_3 \ .
\ea
Using the ansatz (\ref{ansatzC2}) for $C_2$ and the equation (\ref{onshellconstraint}) in the first term, we find
\ba
\label{u(n)onshellsugraaction}
S_{\text{IIB}} =  - \frac{N}{4 \pi k} \int_{\mf{R}^{1,2}} C \we \dd C  + \frac{2 \pi}{4 \kappa_{10} ^2} \int_{\mf{R}^{1,2}  \times D_2 \times  S^5} \Tr b  \we *F_3 \we \d_2 \ .
\ea
Again, the on-shell supergravity action displays both a boundary term and a term evaluated at the center of the disk, that is at the location of the D7-branes. Looking at the boundary term and recalling (\ref{Leffun}), we are already able to state that the holographic theory obtained by imposing Dirichlet boundary condition on $C_2$ is dual to the $U(N)_k$ Chern-Simons, thus confirming the expectation claimed at the beginning of the present subsection. For this to be entirely true, the second (bulk) term in (\ref{u(n)onshellsugraaction}) needs to disappear from the final result of the holographic computation. From the discussion in subsection \ref{secholosu(n)}, we expect it to be canceled by the on-shell D7-branes' action. Using (\ref{relationfortildeCatthetip}) and \eqref{CCTildetip}, we find
\ba
\label{sugractionforun}
S_{\text{IIB}} =  - \frac{N}{4 \pi k}\left. \int_{\mf{R}^{1,2}} C \we \dd C\right|_{r=r_\infty}  + \frac{N}{4 \pi k}\left. \int_{\mf{R}^{1,2} }  C  \we \dd  C\right|_{r=r_0}\ .
\ea
However, the D7-brane action (\ref{d7effectiveaction}) is written in terms of $\tilde C$. Hence, we might as well use again the relation (\ref{CCTildetip}) between the value of $C$ and $\tilde C$ at the tip of the cigar to rewrite (\ref{sugractionforun}) equivalently as
\ba
S_{\text{IIB}} =  - \frac{N}{4 \pi k}\left. \int_{\mf{R}^{1,2}} C \we \dd C\right|_{r=r_\infty}  + \frac{k}{4 \pi N}\left. \int_{\mf{R}^{1,2} }  \tilde C  \we \dd  \tilde C\right|_{r=r_0}\ .
\ea
The sign of the second term may appear as problematic since it seems to prevent the expected cancellation with the D7-branes' effective action (\ref{d7effectiveaction}). However, the holographic theory of this subsection is dual to the $U(N)_k$ theory, which is not level/rank dual to the $U(k)_{-N}$ one, so we do not expect a cancellation with (\ref{d7effectiveaction}). On the contrary, the holographic prescription imposes upon us the cancellation of the bulk term, namely 
\be
S_{\text{IIB}} + S'_{D7} =  - \frac{N}{4\pi k}\left. \int_{\mf{R}^{1,2}} C\wedge \dd C\right|_{r=r_\infty}\ ,
\ee
and therefore demands the theory on the D7-branes to be 
\be
\label{expectedd7action}
 S'_{D7} =  - \frac{N}{4\pi k}\left. \int_{\mf{R}^{1,2}}  C\wedge \dd  C\right|_{r=r_0}\ ,
\ee
which means a $SU(k)_{-N}$ CS theory. To wit, if we change the conditions on the RR fields at $r=r_\infty$, thus performing an $SL(2, \mf{Z})$ operation on the boundary theory, holography instructs us that the same operation has to be applied to the D7-theory that involves the values of the RR fields at $r=r_0$. Indeed, the $SU(k)_{-N}$ CS theory with effective action (\ref{expectedd7action}) is obtained by performing the $S$ operation on the $U(k)_{-N}$ theory displaying effective action (\ref{d7effectiveaction}).
Keeping in mind the caveat mentioned in section \ref{subsecabelianqft}, we can show this by applying the $S$ operation already at the level of the effective action. We thus take (\ref{d7effectiveaction}), replace the background $\tilde C$ with a new dynamical field $c$ and couple the topological symmetry of the latter to a new background field $C$. Hence, we write
\be
\label{d7soperation}
S'_{D7} =   \frac{k}{4\pi N} \left. \int_{\mf{R}^{1,2}} c \wedge \dd c \right|_{r=r_0}  - \frac{1}{2 \pi } \left. \int_{\mf{R}^{1,2}}  C \we \dd c \right|_{r=r_0}\ .
\ee
The equation of motion for $c$ reads
\be
\frac{k}{N} \dd c =  \dd  C  \ .
\ee
Plugging it into the effective action, we indeed find (\ref{expectedd7action}). We also see that on shell, $c$ is really the same field as $\tilde C$, as holography suggests us.

Above, we have hopped between the $C_2$ and the $C_6$ formulations of Type-IIB supergravity without explicitly writing how we perform the electric/magnetic duality. Let us spend some words on this in order to see once again how the theories at the boundary and at the tip are connected. When we want to pass from the supergravity formulation in terms of $C_6$ to that in terms of $C_2$ (and vice-versa), as an intermediate step, we enlarge the theory by considering both $C_2$ and $C_6$ as independent fields. To do this, we need to add to the action the following term:
\be
\label{dualaction}
S_\mathrm{dual} =-  \frac{1}{2 \k_{10} ^2} \int_{\mf{R}^{1,2} \times D_2  \times S^5} F_3\wedge F_7\ .
\ee 
Hence, if we want to work with the theory with $C_2$ ($C_6$) we integrate over $C_6$ ($C_2$). Following the same steps as above, namely  recalling (\ref{formofF3}) and using the ansatz (\ref{ansatzC6}), we have
\ba
S_\mathrm{dual} &=& -  \frac{1}{2 \pi} \left. \int_{\mf{R}^{1,2} } C  \we  \dd \tilde C \right|_{r=r_\infty} +   \frac{1}{2 \pi} \left. \int_{\mf{R}^{1,2} } \Tr b \we \dd \tilde C \right|_{r=r_0} \ .
\ea
Finally, using (\ref{relationforCatthetip}), we find 
\ba
S_\mathrm{dual} &=& -  \frac{1}{2 \pi} \left. \int_{\mf{R}^{1,2} } C  \we  \dd \tilde C \right|_{r=r_\infty} +   \frac{1}{2 \pi} \left. \int_{\mf{R}^{1,2} } C \we \dd \tilde C \right|_{r=r_0} \ .
\ea
We see that the supergravity Lagrangian term that implements the electric/magnetic duality displays a boundary contribution as well as a contribution from the location of the D7-branes. The former is the action term to add to the $SU(N)_k$ effective action at the boundary to find 
the $U(N)_k$ action after integrating over $\tilde C$. The latter is instead a bulk contribution and therefore has to be canceled according to the holographic prescriptions. Indeed, we have shown that we need to add the second term in (\ref{d7soperation}) to the D7-brane action (\ref{d7effectiveaction}) in order to obtain (\ref{expectedd7action}) after integration over $c\equiv \tilde C$.

To summarize, in this subsection we imposed Neumann boundary conditions on $C_6$ and we found that the holographic theory is dual to the $U(N)_k$ CS theory. Starting from the $SU(N)$ theory studied in subsection \ref{secholosu(n)}, turning to Neumann boundary conditions for $C_6$ corresponds to performing the $S$ operation on the holographic theory. On the other hand, in subsection \ref{secholosu(n)} it was shown that holography reproduces the equivalence between the $SU(N)_k$ CS theory and the $U(k)_{-N}$ one, the latter being the theory on the D7-branes in the holographic set-up. In this subsection, we showed that for consistency holography imposes the $S$ operation to be applied to the theory of the D7-branes as well. As a result, even after changing the condition on the fields imposed at the holographic boundary, we find a level/rank duality.

The duality we found in this subsection is $U(N)_k \lra SU(k)_{-N}$. From a quantum field theory point of view, this is the same kind of duality found in subsection \ref{secholosu(n)} since one is obtained from the other by exchanging $N$ and $k$ (and performing a parity transformation). However, since in the holographic context we take $N \gg k$, the two cases are different because the first one establishes the level/rank duality between the strongly-coupled $SU(N)_k$ theory and the weakly-coupled $U(k)_{-N}$ one, whereas the second case does the same for the strongly-coupled $U(N)_k$ theory and the weakly-coupled $SU(k)_{-N}$ one.

Since we learned that holography imposes the same $SL(2,\mf{Z})$ operation at the boundary and on the D7-branes, we are now motivated to look for different boundary conditions leading to the $U(N)_{k, k+Nk'}$ theories at the boundary in order to holographically find level/rank dualities such as (\ref{uulevelrankduality}).

\subsubsection{$U(N)_{k,k+Nk'}$ theory}
\label{seck'}

We would like to now propose how to describe holographically level/rank dualities involving the theories denoted as $U(N)_{k,k+Nk'}$. Such theories are characterized by a one-form symmetry which is $\mf{Z}_{k+Nk'}$. We can see how such a one-form symmetry can arise from \eqref{topologicalsugraterm}. We can simply redefine
\be
C=C'-k'\tilde C\ ,
\ee
so that the topological term coupling to $B_2$ reads
\be
\label{topologicalsugratermk'}
S_{\text{top}} =  - \frac{1}{(2 \pi)^2} \int_{\mf{R}^{1,2} \times r} \pr{ N B_2 \we \dd C' - (k+Nk') B_2 \we \dd \tilde C} \ ,
\ee
and upon fixing $C'$ at the boundary, makes it valued in $\mf{Z}_{k+Nk'}$.

Since we are still fixing a piece of $C_2$, we start from the Type-IIB bulk action formulated in terms of the latter, as in \eqref{iibactionf3} but now with $C_2=2 \pi  (C'-k'\tilde C) \we \dd x^3 / L $.

Focusing for the moment only on the term at the boundary $r=r_\infty$, we get
\ba
S_{\text{IIB,boundary}} =  - \frac{1}{4 \pi } \int_{\mf{R}^{1,2}} C' \we \dd \tilde C  + \frac{k'}{4 \pi } \int_{\mf{R}^{1,2} }  \tilde C  \we \dd  \tilde C\ .
\ea
The relation between $C$ and $\tilde C$ implies that $NC'=(k+Nk')\tilde C$ so that we can rewrite the above as
\ba
S_{\text{IIB,boundary}} =  - \frac{N}{4 \pi (k+Nk')} \int_{\mf{R}^{1,2}} C' \we \dd  C'  - \frac{k'}{4 \pi } \int_{\mf{R}^{1,2} \times r } \dd \tilde C  \we \dd  \tilde C\ .
\label{actionwithk'theta}
\ea
The first term is exactly what we expect for a $U(N)_{k,k+Nk'}$ theory, i.e.~leading to \eqref{topologicalcurrentunkk'}. The second term has been rewritten as an integer $\theta$-term, however in terms of $\tilde C$. At face value, this term should not be even considered, had we kept the bulk IIB supergravity in terms of $C_6$. But we see that it acquires a physical impact if we then revert to the formulation in terms of $C_2$. This is a well known fact, that electric/magnetic duality in the bulk does not commute with integer shifts of the $\theta$ angle, but rather forms the full $SL(2,\mf{Z})$ group out of the combination \cite{Witten:1995gf}. Its holographic interpretation is precisely in terms of the $T$ and $S$ operations on $3d$ field theories \cite{Witten:2003ya}. Indeed, one can obtain the $U(N)_{k,k+Nk'}$ theory by performing $k'$ times the $T$ operation and then the $S$ operation, starting from the $SU(N)_k$ theory. This is exactly what appears to be needed above: start with the $C_6$ formulation appropriate to the $SU(N)_k$ theory, shift the $\theta$ angle by $2\pi k'$ and then go to the $C_2$ formulation.

Now a word about what happens at the tip $r=r_0$, i.e.~on the D7-branes. As we found out in subsection \ref{subsecunk}, the same operations will have to be applied to the $U(k)_{-N}$ theory that lives there to begin with. When $k'=\pm 1$, we end up with the $U(k)_{-N,-N\mp k}$ theories, which are indeed level/rank dual to $U(N)_{k,k\pm N}$ \cite{Hsin:2016blu}. For $|k'|>1$, we do not have a simple description of these theories, but they can still be formally defined (see \cite{Choi:2019eyl}).

We have thus provided a holographic description of this infinite series of level/rank dual theories, where the set-up and the branes involved are still the same, but different boundary conditions are imposed on the RR fields, which, in turn, enforce similar operations on the brane probes at the tip of the bulk geometry.
\section{Discussion}
\label{secdisc}

A summary of our results is the following. We have provided a detailed description of how holography describes level/rank duality, paying attention to the $SU$ or $U$ nature of the gauge groups involved. Such nature must be ascertained on both sides of the duality. At the boundary, this is determined both investigating (holographically) the one-form symmetry, and determining the form of the effective action giving rise to the Hall current. At the other side of the duality, which in our set-up means on the world-volume of the D7-branes at the tip of the cigar, we also determine precisely the nature of the gauge group by the effective action that is required to cancel the contribution from the bulk action. We find perfect agreement with the field theory expectations discussed in subsection \ref{commentonlevelranksimilar}.

Besides corroborating the exact form of the known level/rank dualities, we believe that the most interesting aspect of our results lies in elucidating the role of the bulk $p$-form gauge fields as background fields for symmetry currents. This role is well-known as far as boundary analysis is concerned. What we have shown above is that the same role is assigned to the same fields when probe branes are present in the bulk. In particular, the cancellation between the brane and the bulk contributions at the tip of the cigar can be pictorially interpreted as the bulk transferring the effective action in terms of the background fields from the branes at the tip to the boundary, implementing level/rank and holographic duality at the same time.

Possible extensions of our analysis are to set-ups where flavors are present, which are implemented holographically through probe D5-branes \cite{Jensen:2017xbs} or D7-branes \cite{Hong:2010sb,Argurio:2020her}. It would be interesting to investigate how the dualization of the RR two-form into a six-form and vice-versa affects the couplings in the world-volume of the flavor branes. Another avenue is to investigate non-trivial topologies of the 3$d$ spacetime for which the framing anomaly is relevant, and see how it matches with analogous couplings on the brane actions (similarly as in \cite{Argurio:2018uup}) and in the bulk. Finally, it would be interesting to find a more satisfying understanding of the holographic dual of the $T$ operation, namely a 10$d$ interpretation of the operation of turning on an integer $\theta$ term for $\tilde C$, as discussed below \eqref{actionwithk'theta} (see \cite{Genolini:2021qbi} for a discussion of a similar problem in 11$d$ supergravity).

\vskip 15pt \centerline{\bf Acknowledgments} \vskip 10pt 

\noindent 
We are indebted to Jeremias Aguilera-Damia, Pietro Benetti Genolini, Matteo Bertolini, Francesco Bigazzi, Aldo Cotrone, Carlo Alberto Cremonini, Eduardo Garcia-Valdecasas, Adolfo Guarino, Pierluigi Niro, Colin Sterckx and Luigi Tizzano for helpful discussions. R.A.~is a Research Director of the F.R.S.-FNRS (Belgium), and acknowledges support by IISN-Belgium (convention 4.4503.15) and by the F.R.S.-FNRS under the ``Excellence of Science" EOS be.h project n.~30820817. The work of A.C.~is supported by the ``Fondazione Angelo Della Riccia"  and in part by the ``GGI Boost Fellowship" grant offered by the Galileo Galilei Institute for Theoretical Physics of Florence. R.A.~and A.C.~thank respectively the Galileo Galilei Institute and the Universit\'e Libre de Bruxelles for the kind hospitality during the preparation of this work.


\addcontentsline{toc}{section}{Bibliography}
\bibliography{hololr}
\bibliographystyle{utphys}    

\end{document}